\documentclass[twocolumn,showpacs,preprintnumbers,amsmath,amssymb]{revtex4}
\usepackage{graphicx}
\usepackage{dcolumn}

\def\empile#1\over#2{\mathrel{\mathop{\kern 0pt#1}\limits_{#2}}}

\font\tenmsa=msam10
\font\sevenmsa=msam7
\font\fivemsa=msam5
\font\tenmsb=msbm10
\font\sevenmsb=msbm7
\font\fivemsb=msbm5
\newfam\msafam
\newfam\msbfam
\textfont\msafam=\tenmsa  \scriptfont\msafam=\sevenmsa
\scriptscriptfont\msafam=\fivemsa
\textfont\msbfam=\tenmsb  \scriptfont\msbfam=\sevenmsb
\scriptscriptfont\msbfam=\fivemsb

\catcode`\@=11

\newcount\@tempcntc
\def\@citex[#1]#2{\if@filesw\immediate\write\@auxout{\string\citation{#2}}\fi
  \@tempcnta\z@\@tempcntb\m@ne\def\@citea{}\@cite{%
        \@for\@citeb:=#2\do%
    {\@ifundefined{b@\@citeb}%
        {\@citeo\@tempcntb\m@ne\@citea%
                \def\@citea{,\penalty\@m\ }{\bf ?}\@warning%
                {Citation `\@citeb' on page \thepage \space undefined}}%
        {\setbox\z@\hbox{\global\@tempcntc0\csname b@\@citeb\endcsname\relax}
     \ifnum\@tempcntc=\z@ \@citeo\@tempcntb\m@ne%
       \@citea\def\@citea{,\penalty\@m}%
       \hbox{\csname b@\@citeb\endcsname}%
     \else%
      \advance\@tempcntb\@ne%
      \ifnum\@tempcntb=\@tempcntc%
      \else\advance\@tempcntb\m@ne\@citeo%
      \@tempcnta\@tempcntc\@tempcntb\@tempcntc\fi\fi}}\@citeo}{#1}}%

\def\@citeo{\ifnum\@tempcnta>\@tempcntb\else\@citea
  \def\@citea{,\penalty\@m}%
  \ifnum\@tempcnta=\@tempcntb\the\@tempcnta\else
   {\advance\@tempcnta\@ne\ifnum\@tempcnta=\@tempcntb \else
\def\@citea{--}\fi
    \advance\@tempcnta\m@ne\the\@tempcnta\@citea\the\@tempcntb}\fi\fi}

\catcode`\@=12

\global\mathchardef\lesssim "142E

\newcommand{\slL}{\raise.15ex\hbox{$/$}\kern-.53em\hbox{$L$}}
\newcommand{\slP}{\raise.15ex\hbox{$/$}\kern-.53em\hbox{$P$}}
\newcommand{\slp}{\raise.1ex\hbox{$/$}\kern-.63em\hbox{$p$}}
\newcommand{\slq}{\raise.1ex\hbox{$/$}\kern-.63em\hbox{$q$}}
\newcommand{\slv}{\raise.1ex\hbox{$/$}\kern-.63em\hbox{$v$}}
\newcommand{\slR}{\raise.15ex\hbox{$/$}\kern-.53em\hbox{$R$}}
\newcommand{\slQ}{\raise.15ex\hbox{$/$}\kern-.53em\hbox{$Q$}}
\newcommand{\slK}{\raise.15ex\hbox{$/$}\kern-.53em\hbox{$K$}}
\newcommand{\slk}{\raise.15ex\hbox{$/$}\kern-.53em\hbox{$k$}}
\newcommand{\slSigma}{\raise.15ex\hbox{$/$}\kern-.53em\hbox{$\Sigma$}}
\newcommand{\slcalP}{\raise.15ex\hbox{$/$}\kern-.63em\hbox{$\cal P$}}
\newcommand{\slA}{\raise.15ex\hbox{$/$}\kern-.73em\hbox{$A$}}
\newcommand{\slbfA}{\raise.15ex\hbox{$/$}\kern-.73em\hbox{${\imb A}$}}
\newcommand{\slpartial}{\raise.15ex\hbox{$/$}\kern-.53em\hbox{$\partial$}}

\newcommand{\be}{\begin{equation}}
\newcommand{\ee}{\end{equation}}
\newcommand{\bea}{\begin{eqnarray}}
\newcommand{\ena}{\end{eqnarray}}

\def\build#1\over#2{\mathrel{\mathop{\kern 0pt#1}\limits_{#2}}}

\font\tenimbf=cmmib10 at 10pt
\font\sevenimbf=cmmib10 at 7pt
\font\fiveimbf=cmmib10 at 5pt
\newfam\imbf
\textfont\imbf=\tenimbf
\scriptfont\imbf=\sevenimbf
\scriptscriptfont\imbf=\fiveimbf
\def\imb{\fam\imbf\tenimbf}

\begin{document}

\date{\today}

\title{\bf{Synchrotron radiation in a chromo-magnetic field}}
\author{Alaa Dbeyssi$^{(1)}$, Dima Al Dirani$^{(2)}$,
  H.~Zaraket$^{(2)}$}
\affiliation{
$ (1)$ Institut de Physique Nucl\'eaire d'Orsay,
15 rue G. Cl\'emenceau, 91406 Orsay, France\\
$ (2)$ Laboratoire de Physique de Mat\'eriaux,
Lebanese University, Faculty of Sciences (I),
Hadath, Beirut Lebanon }

\begin{abstract}
We study the generalization of QED synchrotron radiation to the QCD case
with a chromomagnetic field using the Schwinger {\it et al} source method. It is shown
that the QED case can be obtained as a special limit. The comparison with the
path integral approach of Zakharov has shown consistent results.

\end{abstract}
\maketitle
\vskip 4mm

\section{Introduction}
It is believed that during and after relativistic heavy ion collisions
strong chromomagnetic field will form that can be treated as a classical
background. Numerical solutions \cite{FriesK,LappiM} indicate that ''just"
after collision transverse color electric and color magnetic fields
change suddenly from being transverse in the initial state, in the so-called Color-Glass-Condensate state,
to being longitudinal. The latter are called glasma flux tubes.
Transverse fields then rise, and at some stage after the collision
the transverse and longitudinal
components of color electric and color magnetic fields reach a
''steady" comparable values.  In this rich environment a fast
parton escaping from the collision would feel the effect of such a
field. Synchrotron and \v{C}erenkov effects are important physical phenomena
to be studied, either as an energy loss mechanism or as a coherent
gluon radiation process. In this context the \v{C}erenkov radiation was
 considered  in \cite{Dremin}. Syncrotron radiation was analyzed
 for a longitudinal field in \cite{ShuryaZ}, and a transverse field in
 \cite{Zak1}. We focus in this paper on the synchrotron radiation by a moving quark in a
longitudinal chromomagnetic field; this should be distinguished from the
case studied in \cite{Tuchin}.  The author of \cite{Tuchin} has considered
the motion of fast fermion in an "electromagnetic" magnetic field
not a chromomagnetic field.

The synchrotron radiation of a photon from a relativistic electron
was well studied in the last century \cite{Soko,Schwinger}.
Generalizing the QED study to the QCD case has been done in
\cite{ShuryaZ} for a chromomagnetic field. It would have been the
"end of a new chapter" but in \cite{Zak1} the
path integral approach was used to derive the synchrotron radiation of gluons
by a parton in a transverse chromomagnetic field.  The results found in
\cite{Zak1} seem to be radically different from \cite{ShuryaZ}; it was
also argued that the
QED case cannot be obtained from the QCD generalization of \cite{ShuryaZ}. This
motivates reconsidering the QCD generalization of the QED operator/source
method used in \cite{SchwinET}. We show that the QED case is
simply obtained from our QCD generalization. After clarifying the differences
in the field configurations considered in \cite{ShuryaZ} and \cite{Zak1}, it
is possible to select the closest possible configuration to \cite{Zak1} and
make a comparison between our results and that of \cite{Zak1}. We show that
very similar expressions are found. We analyze some of the
possible inconsistencies that led to the erroneous results of \cite{ShuryaZ}.

The radiation in external magnetic fields is
applicable in all Lorentz frames where $H^2 - {\cal E}^2 > 0$, where
${\cal E}$ is the electric field. Hence
we consider a weak field approximation to avoid
the ''Klein catastrophe'', that is, spontaneous pair creation by an electric
field. Hence for $|F_{\mu\nu}F^{\mu\nu}|^{1/2}\ll m^2/g$, pair creation effects are negligible.
Besides, for the semiclassical approximation to hold we have to have $g^3(F/m^2)(E/m)\ll 1$; having large
quark energy the only configuration to have a semiclassical approximation is in the weak field
regime, {\it i.e.}, $H\ll m^2/g$. Therefore we consider the case where  $H\ll m^2/g$.

The paper is organized as follows. The coupling of quarks and
gluons to the background field is first represented in terms of
color charges. The Schwinger {\it et al} method  is then presented, the gluon synchrotron
rate is calculated, and finally we compare the obtained results with
different results in literature.


\section{Charge "representation"}
The main difference between the QED and QCD synchrotron is the
interaction between the emitted gluon and the background chromomagnetic field.
The non-Abelian generalization of the QED synchrotron radiation can be
easily obtained if we characterize the coupling between
gluons and the chromomagnetic background field through a fictive color
charge. The fictive charge can be explicitly obtained by starting from the
QCD Lagrangian, decomposing the gauge field into quantum fluctuations and
a classical background field, then regrouping terms that couple to a given
background color index. The coupling constant $g$ is then found to be multiplied
by a number which can then be interpreted as a color charge. The more
formal approach is to use the Cartan subgroup of SU(3), then quarks and gluons
can be characterized by two color charges. SU(3) is a unimodular
group of 3$\times$3 Hermitian linearly independent matrices of determinant equal
to 1. SU(3) has $3^2-1=8$ generators $\lambda_i$ ($i=1\cdots 8$). The
maximum number of commuting generators for SU(3) is $2$. In the
Gell-Mann fundamental representation $\lambda_3$ and $\lambda_8$
commute hence ${\cal H}_1=\lambda_3/2$ and ${\cal H}_2=\lambda_8/2$ form the
Cartan subgroup
(they can be seen as the analogue of $J_3$ for SU(2)) and are already
diagonal. Define the following combinations of the remaining six generators:
$$ E_{\pm \alpha}=\lambda_1\pm i\lambda_2 \quad E_{\pm \beta}=\lambda_6\pm i\lambda_7\;\; {\rm and}\; E_{\pm \gamma}=\lambda_4 \pm i\lambda_5$$
these new matrices will play the role of raising and lowering operators, they satisfy
$$[{\cal H}_i,E_{\xi}]=\xi({\cal H}_i)E_{\xi}\; . $$
Hence the $E_{\xi}$ are eigenvectors of the
$ad({\cal H}_1)$ and $ad({\cal H}_2)$. The weight $\xi$ of the
adjoint representation are called the roots. They are the color charge of the
gluon.   A simple calculation, for the Gell-Mann representation, gives the
gluon charges ($\xi({\cal H}_1),\xi({\cal H}_2)$ or $(Q_3,Q_8)$)
$$ (1,0)\;,\;(-1,0)\;,\;(-\frac{1}{2},\frac{\sqrt{3}}{2})\; ,\; (\frac{1}{2},-\frac{\sqrt{3}}{2})\; ,$$
$$(\frac{1}{2},\frac{\sqrt{3}}{2})\; ,\;(\frac{1}{2},\frac{\sqrt{3}}{2})\; ,\;(-\frac{1}{2},-\frac{\sqrt{3}}{2})\;,(0,0)\; .$$
Quarks are found in the fundamental representation, so the eigenvalues of
${\cal H}_i$ give the color charges
of quarks. So quark "states" will be common eigenstates of
${\cal H}_1,{\cal H}_2 $ with eigenvalues $(q_3,q_8)$):
$$(\frac{1}{2},\frac{\sqrt{3}}{6})\; ,\;  (-\frac{1}{2},\frac{\sqrt{3}}{6})\; ,\; (0,-\frac{\sqrt{3}}{3})\; .$$
The color charge
will always be multiplied by the strong interaction
coupling constant $g$, so it is possible to absorb the "charge" in the
coupling and define a new coupling $g^i_A=gQ_i$ for gluons and
$g^i_q=gq_i$. To simplify our calculation we will consider a
chromomagnetic field with a given color index, either $3$ or $8$ but not
combination of both color indices.

\section{The mass operator}
In the Schwinger {\it et al}  method, the emission process can be seen through
a modification of the Dirac equation of a spin $1/2$ particle of mass $m$ in
the presence of an external field $H$. Besides the usual modification of the
momentum operator (a tree level modification) to include the external field
effect ($i\partial_\mu\rightarrow i\partial -qA_\mu=\Pi_\mu$), the emission
process  can be seen as a modification of the mass by an operator
$M$ called the mass operator:
\begin{equation}
\left(\gamma_\mu\Pi^\mu+m+M\right)\Psi=0\; .
\end{equation}
So the mass operator is always acting on a physical spin $1/2$ real
particle state. As we will see later, special attention should be paid in
some cases to its correct dispersion relation in the presence of an external field.

The total decay rate $\Gamma$ of a particle of mass $m$ and energy $E$
can be obtained from the mass operator $M$ by an
"optical theorem"-like relation
\begin{equation}
\Gamma=-2\frac{m}{E}{\rm Im}\; M
\label{eq:mass}
\end{equation}
where $M$ is obtained from the one-loop diagram with full quark and gluon
propagators in the presence of the field $H$. So the final expression
will be valid to all orders in $gH$ but not to all orders in $\alpha_s$.
The decay rate in eq.~(\ref{eq:mass}) is a number. $M$ is understood as
an average of the mass operator over different spin states:
$$ M\rightarrow \frac{\bar{u}\hat{M} u}{\bar{u}u}\; .$$

However we are usually interested in finding the probability $P$ of emitting
a gluon with a given energy $\omega$. In such a case the radiation power
for a given energy is related to ${\rm Im}\; M$ by the integral equation
\begin{equation}
-\frac{1}{E}{\rm Im}\; M=\int\frac{d\omega}{\omega}P(\omega)\; .
\end{equation}
Hence to get $P(\omega)$ we have to write $M$ in an integral form as we will
see later.

\section{Effective propagators}
Having a strong magnetic field, it is necessary to evaluate the mass operator
to all orders in the field $H$. We consider an "external" (classical)
chromomagnetic field along the longitudinal direction (the third axis or the
$z$ axis):
$$H_\mu^a=\delta_{\mu 3}\delta^{ai} h$$
where the color index $i$ is either $3$ or $8$. This field will lead to
gauge field $A$ that is linearly dependent on position.

It is necessary to use effective quark and gluon propagators in the presence
of $H$.

The quark propagator in a field $H$ can be written in a Fock-Schwinger proper time
form as
\begin{equation}
S(x,x^\prime)=\phi_q(x,x^\prime) \int \frac{d^4p}{(2\pi)^4}\exp[ip(x-x^\prime)]S(p)
\end{equation}
where, the non-translation invariant, gauge dependent factor $\phi$ is a
Bohm-Aharonov-like phase
$$\phi_q(x,x^\prime)=\exp[igq\int_x^{x^\prime} d\chi_\mu A^\mu(\chi)] $$
The "momentum" space quark propagator is
\begin{eqnarray}
S(p)=i\int_0^\infty \frac{ds_2}{\cos z} \exp\{-is_2[m^2-p_\parallel^2+\frac{\tan z}{z}p_\perp^2 ]\}
\nonumber\\
\times\{m e^{iz\Sigma_3}+e^{iz\Sigma_3}(\vec{\gamma}\cdot p)_\parallel-\frac{1}{\cos z}(\gamma\cdot p)_\parallel\}
\end{eqnarray}
We have explicitly used the longitudinal and transverse decomposition appropriate
for the choice of the field: $(\vec{a}\cdot \vec{b})_\perp=a_1b_1+a_2b_2$,
$(a\cdot b)_\parallel=a_ob_o-a_3b_3$. The parameter $z$ depends on the
color charge of the considered quark $z=gqhs_2$. The matrix $\Sigma_3$ is the
usual ($z$-)spin-projection. Graphically, in Feynman diagrams, the effective
 quark will have a blob to distinguish it from bare propagators.

The nonzero coupling between the gluon and the background leads
to a modification of the bare propagator. The gluon
effective propagator can be written in a form similar to the quark propagator
\begin{equation}
G_{\mu\nu}(x,x^\prime)=\phi_{_{Q}}(x,x^\prime) \int \frac{d^4p}{(2\pi)^4}\exp[ip(x-x^\prime)]G_{\mu\nu}(p)
\end{equation}
where the gluon phase factor is again
 $$\phi_{_{Q}}(x,x^\prime)=\exp[igQ\int_x^{x^\prime} d\chi_\mu A^\mu(\chi)] $$
 and the "momentum" representation gluon propagator is
\begin{equation}
G_{\mu\nu}(p)=i\int_0^\infty \frac{ds_1}{\cos y} \exp\{-is_1[-p_\parallel^2+\frac{\tan y}{y}p_\perp^2 ]\}E_{\mu\nu}
\end{equation}
where $y=gQhs_1$, and the tensor $E_{\mu\nu}$ can be written (formally) in a
compact form in terms of the tensor $F_{\mu\nu}$ as
 \begin{equation}
 E_{\mu\nu}=[\exp(2gQFs_1)]_{\mu\nu}
 \end{equation}
 A more practical form is
\begin{equation}
 E_{\mu\nu}=g_{\mu\nu}^\parallel+g_{\mu\nu}^\perp\cos(2y)-A_{\mu\nu}\sin(2y)\; .
 \end{equation}
 The metric tensors $g_{\mu\nu}^{\parallel,\perp}$ stand for the longitudinal and
 transverse spaces metric. The antisymmetric tensor $A$ is
 $$ A_{\mu\nu}=\delta_{\mu 1}\delta_{\nu 2}-\delta_{\mu 2}\delta_{\nu 1}\; .$$
 It is easy to verify that in the limit of the vanishing magnetic field $E_{\mu\nu}\rightarrow g_{\mu\nu}$.
 \section{One-loop contribution}
\begin{figure}
\centerline{\resizebox*{!}{3cm}{\includegraphics{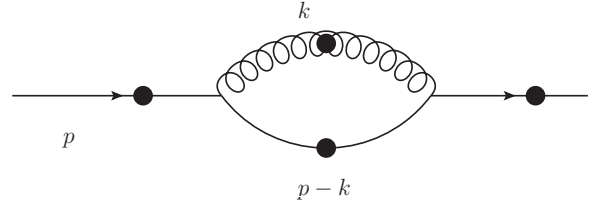}}}
\caption{\label{fig:1loop} The one-loop diagram contributing to
  the mass operator in momentum space. Gluon and quark propagators are effective
  propagators in the presence of an external chromomagnetic field. }
\end{figure}
The order $\alpha_s$ contribution to the ''mass operator'' is given by the
diagram shown  in Fig.~\ref{fig:1loop}. The coordinate space $M$ is given then by
\begin{equation}
M(x,x^\prime)=\phi_q(x,x^\prime)\int\frac{d^4p}{(2\pi)^4}\exp[ip(x-x^\prime)]M(p)
\end{equation}
where we have used the relation between the external quark charge $q$ and the
the charges of the internal quark $q^\prime$ and the gluon $Q$:
$q=q^\prime+Q$ to write
$$\phi_q(x,x^\prime)=\phi_{q^\prime}(x,x^\prime)\phi_{_{Q}}(x,x^\prime)\; . $$
The "momentum" representation mass operator is
\begin{equation}
M(p)=-g^2(T^A)_{ik}(T^A)_{ki}\int \frac{d^4k}{(2\pi)^4} G_{\mu\nu}(k)\gamma^\mu S(p-k)\gamma^\nu\; .
\end{equation}
To get the un-integrated emission probability $P(\omega)$, we use the insertion
used in \cite{SchwinET}:
$$1=\int\limits_{-\infty}^\infty d\omega\int\limits_{-\infty}^\infty\frac{d\lambda}{2\pi}\exp\{-i\lambda(\omega-k_o)\}\; .$$
From now on our procedure will follow the steps of
\cite{SchwinET}. To avoid hindering the main result of this paper by technical details we present
the main steps and most of the derivation Appendix \ref{app: a}.

The exact expression of the one-loop momentum representation
$M(p)$, after $k$ and $\lambda$ integration, is
\begin{eqnarray}
M(p)=-2\frac{g^2}{(4\pi)^2}(T^A)_{ik}(T^A)_{ki}\int\limits_{-\infty}^{\infty} d\omega\int\limits_0^\infty ds \left(\frac{i}{\pi s}\right)^{\frac{1}{2}}\nonumber\\
\int\limits_0^1\frac{du}{\delta^\prime\cos y\cos z}\exp(is\Phi)\left\{ me^{iz\Sigma_3}+\frac{(1-u)\tan y}{y\delta^\prime \cos z} (\vec{\gamma}\cdot\vec{p})_\perp\right.\nonumber\\
\left.\hspace{-2mm}+\left[m-(p_o-\omega)\gamma_o+p_z(1-u)\gamma_3\right]e^{-i(z-2y)\Sigma_3}\right\}
\label{eq: M}
\end{eqnarray}
Few definitions are made: $y=(1-u)sgQh$; $z=usgq^\prime h$
$$ \delta^\prime = (1-u)\frac{\tan y}{y}+u\frac{\tan z}{z}$$
and the phase factor
$$\Phi=-m^2+u(1-u)p_\parallel^2 -u(1-u)\frac{\tan z}{\delta^\prime z}\frac{\tan y}{y}+(up_o-\omega)^2\; .$$
Note that $M(p)$ at this stage is still in a matrix form.

\section{Fourier transform: dispersion relation}

The expression of $M(p)$ found in the previous section [Eq. (\ref{eq: M})] has
to be Fourier transformed and then multiplied by the gauge field depending
phase $\phi_q(x,x^\prime)$, with the external quark charge $q$, to give
$M(x,x^\prime)$. This can be done using the relations derived by Tsai in \cite{Tsai}.  For instance
\begin{eqnarray}
\phi(x,x^\prime)\int\frac{d^4 p}{(2\pi)^4} e^{ip(x-x^\prime)}e^{iA_op_o^2}e^{iA_3p_3^2}e^{-iBp_\perp^2}\nonumber\\
=\cos Z\langle x|e^{iA_o\Pi_o^2}e^{iA_3\Pi_3^2}e^{i{\cal S}\Pi_\perp^2}|x^\prime\rangle
\label{eq: tsai}
\end{eqnarray}
The $\Pi$'s are the generalized momentum operators in the presence of a
magnetic field. This relation can be easily derived, as shown in Appendix~\ref{app: b}, if one notices that the
right-hand side can be decomposed into three quantum non-relativistic
propagators: a free particle of some fictive mass related to $A_o$ living
in the "$o$" one-dimensional space,
a free particle living in the "$3$" one dimensional space, and finally a
particle living in the transverse plane and under the action of an external
magnetic field $H$ perpendicular to the plane. The three propagators are
exactly known which give the above relation with $Z=gq{\cal S}$ and
$${\cal S}=\frac{1}{gqh}\tan^{-1}(gqhB) $$
Similar relations are used for the product of exponential factors and $p_\mu$
to the left of Eq.~(\ref{eq: tsai}).
The mass operator can then be obtained. This mass operator will be acting on
physical quark states satisfying $(\Pi_\mu\gamma^\mu+m)\psi=0$.  Hence it is possible to do the following replacements
$$\hat{\Pi}_\perp^2 \rightarrow \hat{\Pi}_\parallel^2+gqh\Sigma_3-m^2 \; ;\quad \hat{\Pi}_\parallel\rightarrow p_\parallel;,$$
where $p_\parallel$ now represents the eigenvalues of $\hat{\Pi}_\parallel$ which is not
affected by the longitudinal chromomagnetic field.

Applying Tsai's relations and the external quark equation of motion to
Eq.~(\ref{eq: M}) the mass operator will be
\begin{eqnarray}
\hat{M}=-2\frac{g^2}{(4\pi)^2}(T^A)_{ik}(T^A)_{ki}\int d\omega\int ds \left(\frac{i}{\pi s}\right)^{\frac{1}{2}}\nonumber\\
\int du\frac{\cos Z}{\delta^\prime\cos y\cos z}\exp(is\Phi^\prime)
\left\{ me^{i(z-Z)\Sigma_3} \right.\nonumber\\
\left. +\cos Z\frac{(1-u)\tan y}{y\delta^\prime \cos z} [(\gamma\cdot p)_\parallel-m] \right. \nonumber \\
\left.\!\!\!\!\!\!\!\!+\left[m-(p_o-\omega)\gamma_o+p_z(1-u)\gamma_3\right]e^{-i(z+Z-2y)\Sigma_3}\right\}
\label{eq: Mx}
\end{eqnarray}
where $Z=gq{\cal S}$, and
$$ {\cal S}=\frac{1}{gqh}\tan^{-1}\left[su(1-u)\frac{\tan z}{\delta^\prime}\frac{\tan y}{y}\right]$$
defined as mentioned before from the Fourier transform. The new
phase factor, which is a scalar now, is
\begin{equation}
\Phi^\prime=u(p_o-\omega)^2-(1-u)(up_3-\omega^2)-\frac{\cal S}{s}(p_o^2-p_3^2-m^2)\; .
\end{equation}

\section{Gamma substitutions}
The mass operator has a spinorial/matrix structure. It should be mentioned
that we have used the
exact dispersion relation for the replacement of $\Pi_\perp$ in the phase
factor, without solving the equation of motion.  However, the Dirac structure
in front of the exponential factor in (\ref{eq: Mx}) depends on the gamma
matrices in a nontrivial way. We use the same approximation done in the
literature, which is the weak point of the method: the spinor structure of the incident quark is the same as that of
the free particle, {\it i.e.}, the Dirac spinor $u_s(p)$. So sandwiching the
mass operator between $u$ and $\bar{u}$, the gamma matrices are replaced by
scalar quantities. Simple matrix calculation leads to the following replacement
rules: $\gamma_o\rightarrow p_o/m$, $\gamma_3\rightarrow p_3/m$,
$\Sigma_3\rightarrow \eta \frac{1}{m}\left(p_o -\frac{p_z^2}{p_o+m}\right)$; $\gamma_o\Sigma_3\rightarrow\eta(1+\frac{p_3^2}{m(p_o+m)})$; $\gamma_3\Sigma_3\rightarrow\eta\frac{p_3}{m}$.

The $\eta=\pm$ are for the different helicities of the quark. The gamma replacement leads to a scalar mass operator that can be used to
obtain $P(\omega)$,
\begin{eqnarray}
M=-2\frac{g^2}{(4\pi)^2}(T^A)_{ik}(T^A)_{ki}\int d\omega\int ds \left(\frac{i}{\pi s}\right)^{\frac{1}{2}}\int du\nonumber\\
\frac{\cos Z}{\delta^\prime\cos y\cos z}\exp(is\Phi^\prime)\left\{ m(\cos(z-Z)+\cos(Z+z-2y))\right.\nonumber \\
 -i\eta\left(p_o-\frac{p_z^2}{p_o+m}\right)(\sin(Z-z)+\sin(Z+z-2y)) \nonumber\\
+\cos Z\frac{(1-u)\tan y}{y\delta^\prime \cos z}\left(\frac{p_o^2-p_3^2-m^2}{m}\right)
\nonumber\\
-(p_o-\omega)\frac{p_o}{m}\cos(Z+z-2y)+\frac{p_z^2}{m}(1-u)\cos(Z+z-2y)\nonumber\\
+i\eta(p_o-\omega)\left(1+\frac{p_3^2}{m(p_o+m)}\right)\sin(Z+z-2y)
\nonumber\\
\left.-i\eta (1-u)\frac{p_z^2}{m^2}\left(p_o-\frac{p_z^2}{p_o+m}\right)\sin(Z+z-2y)\right\}
\label{eq: Mxgamma}
\end{eqnarray}
Note that this expression has helicity information that can be explored in
a way similar to \cite{Tuchin}.

\subsection{The Abelian limit}
The results found in \cite{SchwinET} can be easily obtained from our expression
in Eq.~(\ref{eq: Mxgamma}). The non-Abelian character can be waived if we set the
gluon charge $Q$ to zero hence $y=0$; remove the color factor $T^AT^A$ in $M$; and
consider the motion of the incident quark in the transverse plane, {\it i.e.}, take
$p_3=0$. Besides, we have to set the index of refraction $n=1$ in
\cite{SchwinET}; hence the parameters $\delta$, $E$,
$\widetilde\beta$, $\delta^\prime$ of
\cite{SchwinET} will be in our notation (for $y=0=p_3$) $\delta=1$,
$E\rightarrow p_o$,  $\delta^\prime\rightarrow \delta^\prime $, $\widetilde\beta\rightarrow Z$.

The above simplifications lead to
\begin{eqnarray}
M_{abelian}=2\frac{g^2}{(4\pi)^2}\int d\omega\int ds \left(\frac{i}{\pi s}\right)^{\frac{1}{2}}\int du\exp(is\Phi^\prime)\frac{m}{\Delta^{1/2}}\nonumber\\
\left\{ \cos(Z-z) \right.
-i\eta\frac{p_o}{m}\sin(Z-z)\nonumber\\
+\frac{\omega}{m}\frac{p_o}{m}\cos(Z+z) -i\eta\frac{\omega}{m}\sin(Z+z)\nonumber\\
\left.  +\frac{p_o^2-m^2}{m^2}\left[\frac{1-u}{\Delta^{1/2}}-\cos(Z+z)\right]\!\!\right\}
\label{eq: Schwin}
\end{eqnarray}
whereas in \cite{SchwinET} $\Delta^{1/2}=(\delta^\prime\cos z)/\cos Z$.
Eq.~(\ref{eq: Schwin}) coincides with equation (3.39~a) of \cite{SchwinET}.

\subsection{Small $s$ limit}
By analyzing Eq.~(\ref{eq: Mxgamma}) it is clear that the $s$-integration
is dominated by small a $s$ region. The small $s$ expansion of the phase factor
$\Phi^\prime$ gives
\begin{eqnarray}
\Phi^\prime \approx (up_o-\omega)^2 - m^2u^2 \nonumber\\
-\left[gq^\prime u h
-gQh(1-u)\right]^2\frac{s^2}{3}\left[u^2(1-u)^2(p_o^2-p_3^2)-m^2\right]
\end{eqnarray}
The $u$-integration can then be approximated by a Gaussian integration:
\begin{equation}
\int_0^1 du f(u)e^{is(up_o-u)^2}\approx f(u\approx \frac{\omega}{p_o})\frac{1}{2ip_o}\sqrt{\frac{\pi}{is}}
\end{equation}
In this approximation the variable $u$ is replaced by the
gluon energy fraction $x=\omega/p_o$. After the $u$-integration the phase
factor will be
\begin{equation}
s\Phi^\prime \rightarrow -m^2sx^2-\frac{\mathbf{f}^2}{3}s^3 \left[x^2(1-x)^2(p_o^2-p_3^2)-m^2\right]
\label{eq: phase}
\end{equation}
where we have defined the vector
$$ \mathbf{f}=gq^\prime x\mathbf{H}-gQ (1-x)\mathbf{H}$$
which represents a fictive magnetic force on an intermediate gluon-quark system.
\subsection{Zakharov method}
Before proceeding to an explicit comparison of our result with the one
found in \cite{Zak1}; we present Zakharov's method in its simplest "display."

Consider first a fast scalar particle whose first quantized wave function
satisfying the Klein-Gordon equation
\begin{equation}
(\frac{\partial^2}{\partial t^2}-\frac{\partial^2}{\partial\mathbf{r}^2})\psi=m^2\psi\; .
\end{equation}
For a fast particle moving along the $z$-axis with large energy $E$, the wave function can
be written as
\begin{equation}
\psi(t,\mathbf{r})=\exp\left(-iE\xi\right)\phi(z,\xi,\mathbf{\rho})
\end{equation}
where $\xi=t-z$, $\mathbf{r}=(z,\mathbf{\rho})$. For fixed $\xi$ the "transverse" wave function
$\phi$ satisfies the equation
\begin{equation}
i\frac{\partial\phi}{\partial z}=\left(\frac{\mathbf{p}_\perp^2+m^2}{2E}\right)\phi+\frac{\partial^2\phi}{\partial z^2}
\end{equation}

Besides, in the low mass limit or for $m\ll E$ the above equation simplifies to
\begin{equation}
i\frac{\partial\phi}{\partial z}=\left(\frac{\mathbf{p}_\perp^2+m^2}{2E}\right)\phi\; .
\end{equation}
Which is a two-dimensional Schr\"odinger equation with a mass $E$. The solution to the above equation can be easily
obtained,
\begin{equation}
\phi_{p_z,p_\perp}(z,{\bf \rho})=e^{i{\mathbf{p}_\perp}\cdot{\bf \rho}}e^{-i\int\limits_0^z\left(\frac{\mathbf{p}_\perp^2+m^2}{2p_z}\right)dz}\; ,
\end{equation}
which can be interpreted as a plane wave. The $\xi$ dependence emerges only via boundary
conditions for the transverse wave function. So we have an evolution equation along each line $\xi={\rm const}$.

If we consider now a parton of color charge $Q$ in an external field
$H$, represented by a gauge field $A_\mu$, the
equation of motion becomes
\begin{equation}
i\frac{\partial\phi}{\partial z}=\left\{\frac{(\mathbf{p}-gQ\mathbf{A})_\perp^2+m^2}{2E}+gQ(A^o-A^3)\right\}\phi\; .
\end{equation}
The field configuration considered in \cite{Zak1} is such that
$\vec{A}=0=A^o$ and $A_3=[\mathbf{H}\times \mathbf{\rho}]_3$; hence the magnetic field is taken to be in the transverse plane. The wave function is then simply
found to be
\begin{equation}
\phi_{p_z,p_\perp}(z,{\bf \rho})=e^{i{\vec{p}_\perp}\cdot{\bf \rho}}e^{-i\int\limits_0^z\left(\frac{\mathbf{p}_\perp^2+m^2}{2p_z}\right)dz}\; ,
\end{equation}
But now the transverse momentum is a solution of the "classical equation of motion"
$$\frac{d\mathbf{p}_\perp}{dz}=\mathbf{F}\; . $$
This is applied to the incident quark as well as to the outgoing gluon and quark
 \footnote{If we wish to study the longitudinal magnetic field we have to consider keeping
$\mathbf{A}_\perp$ which will introduce quadratic terms in the two-dimensional 
Hamiltonian which will complicate the procedure.}.

The final result of \cite{Zak1} can be written as
\begin{eqnarray}
\frac{dP}{dLdx}=\frac{i\mu}{2\pi}\int\limits_\infty^{-\infty}\frac{d\tau}{\tau}\left[\frac{g_1}{\mu^2}\left(\epsilon^2+\frac{\mathbf{f^2\tau^2}}{2}\right)-g_2\right]\nonumber\\
\exp\left\{-i\left[\frac{\epsilon^2\tau}{2\mu}+\frac{\mathbf{f}^2\tau^3}{24\mu}\right]\right\}
\label{eq: Zak}
\end{eqnarray}
where $\mu=Ex(1-x)$ (an effective mass), $g_1=C\alpha_s(1-x+x^2)/x$ (non-spin-flip 
vertex factor), $g_2=C\alpha_sm_q^2x^3/2\mu^2$ (spin-flip vertex factor),
$\epsilon=m_q^2x^2+m_g^2(1-x)^2$, and as in our approach the fictive force
of a quark-gluon system is $\mathbf{f}=x\mathbf{F}_{q^\prime}-(1-x)\mathbf{F}_g$.

Comparing a particle propagation in a longitudinal field to that of a transverse
field configuration is not intuitive. However, in an infinite (nonrealistic)
medium, and for a static (non-propagating) magnetic field, and with a vanishing
electric field, it is the propagation direction of the incident parton which gives
the meaning of the longitudinal/transverse directions. So if in our approach
we take $p_z=0$ we will be the closest to the configuration of
\cite{Zak1}, where the transverse direction is mapped
into the longitudinal direction in the following ways:
\begin{itemize}
\item Present configuration: a particle moving in the transverse plane under the
action of a longitudinal field.
\item Zakharov's configuration: a fast particle moving along the longitudinal
direction and a transverse magnetic field.
\end{itemize}
 It is closest but not the same, since in \cite{Zak1}, transverse momenta are
 not set to zero.

So setting $p_z=0$, $p_o=E$ and $ s=\tau/(2\mu)$ in Eq.~(\ref{eq: phase}) we
obtain the same phase given in Eq.~(\ref{eq: Zak}) for a zero gluon mass.
However, not surprisingly, the prefactor of our approximated expression does
not match that of Eq.~(\ref{eq: Zak}).

\subsection{Shuryak et al}
It is now clear that our agreement with \cite{Zak1} leads to the
conclusion that the results of \cite{ShuryaZ} are ''defective". As the
author of \cite{ShuryaZ} did not give enough details about their
approximation method it is hard to trace the exact source of
inconsistency  in \cite{ShuryaZ}. However we expect that their asymmetric
treatment of the quark and gluon propagators leads to a nonsystematic approximation/expansion.
For instance, the important Bohm-Aharonov phase, which plays an important role in our derivation, was absorbed
in an unjustified way. Consequently the
fictive force $\mathbf{f}$ was badly approximated.


\section{Conclusions}
The generalization of QED synchrotron radiation to the QCD case was done
with the minimal possible complications. It is shown that the QED case can
be obtained as a special limit. The comparison with the path integral approach
of Zakharov has shown consistent results. This will
hopefully "solve" the debate about gluon synchrotron radiation.

It is possible now to extend our method to include nonzero gluon mass
(thermal mass) and combinations of longitudinal and transverse
chromoelectric and chromomagnetic fields with an arbitrary color index. This is
under current investigation.

\section*{Acknowledgments}
The work of H.Z. was supported by the Lebanese University. He would also like
to thank his colleagues at the Laboratoire d'Annecy le Vieux de Physique
Th\'eorique for their hospitality. This paper was initially planned
and discussed with P. Aurenche, we would like to thank him sincerely.
We would also like to thank B.G. Zakharov for useful discussions.

\appendix
\section{Schwinger et al procedure}
\label{app: a}
Our derivation follows the recipe of \cite{SchwinET}. We present in what follows the main steps
needed to get equation (\ref{eq: M}).

The first step to follow is the change of variables: $s_1=(1-u)s$, $s_2=us$,
where $0\le u\le 1$, and $s$ is positive. The $\lambda$ and $k$ integration is based on
Gaussian integration and saddle point approximation. Hence the exponential factor is factored out in
the expression of $M$; this can be written as
\begin{equation}
M(p)=\int\limits_0^1 du \int d^4k\int\limits_{-\infty}^\infty d\lambda d\omega \int\limits_0^\infty ds M_{_{\rm Dirac}} \exp \left(is(\chi -\frac{\lambda \omega}{s})\right)
\end{equation}
where $M_{_{\rm Dirac}}$ contains the Dirac structure of the self-energy diagram and all the non-exponential terms. The function $\chi$ can be
split into a momentum dependent part $\chi_1$ and a $\lambda$ dependent function $\chi_o$, $\chi=\chi_o+\chi_1$, such that
$$\chi_1=\left(k_o-(up_o-\frac{\lambda}{2s})\right)^2-(k_3-up_3)^2-\delta^\prime\left(\vec{k}_{\perp}-\frac{u\tan z}{z \delta^\prime}\vec{p}_{\perp}\right)^2$$
while
$$ \chi_o=-\frac{1}{s^2}\left(\frac{\lambda^2}{4}-up_o\lambda s\right)+\Phi$$
where $\delta^\prime$ and $\Phi$ are as defined before.

We perform now the $k$ integration using the Gaussian form(s)
\begin{eqnarray}
&&\int \frac{d^4k}{(2\pi)^4}\exp(is\chi_1)\left[1,k_o,k_3,\vec{k}_\perp\right]=\nonumber\\
&&\frac{i}{\delta^\prime(4\pi)^2s^2}\left[1,up_o-\frac{\lambda}{2s},up_3,\frac{u\tan z}{\delta^\prime z}\vec{p}_{\perp}\right]
\end{eqnarray}
For the $\lambda$ integration
\begin{eqnarray}
&&\int \frac{d\lambda}{2\pi} \exp(is\chi_o-i\lambda \omega)[1,\lambda]=\nonumber\\
&&\left(\frac{s}{i\pi}\right)^{1/2}e^{is(\Phi+(up_o-\omega)^2)}[1,2s(up_o-\omega)]
\end{eqnarray}
The remaining, lengthy but straightforward, step needed to get equation (\ref{eq: M}) is the
Dirac structure simplification/contractions.

\section{Tsai transformation to coordinate space}
\label{app: b}
We give a simple derivation for equation~(\ref{eq: tsai}) which is different from the initial method
of Tsai \cite{Tsai}.

Consider a non-relativistic particle of mass $m$ and charge $g$ in a uniform magnetic
field $H$. The general expression of the non-relativistic propagator in coordinate space is
\begin{eqnarray}
K(\vec{r},\tau,\vec{r}_o,0)=\langle \vec{r}|\exp\left(-i{\cal H}\frac{\tau}{\hbar}\right)|\vec{r}_o\rangle\nonumber\\
=\left(\frac{m}{2i\pi\hbar s}\right)^{3/2}\frac{gH\tau/2m}{\sin(gH\tau)/2m}\exp\left(\frac{iS_{\rm cl}}{\hbar}\right)
\end{eqnarray}
where ${\cal H}$ is the particle Hamiltonian
$${\cal H}= \frac{\Pi_\perp^2}{2m}-\frac{\Pi_\parallel^2}{2m} $$
with $\Pi$ the generalized momentum.  The classical action $S_{\rm cl}$ in the presence of
a magnetic field is
\begin{eqnarray}
S_{\rm cl}=\frac{gH}{2}\{\frac{1}{2}\left[(x-x_o)^2+(y-y_o)^2\right]\cot(gH\tau/2m)\nonumber\\
 +(x_oy-y_ox)\}+\frac{m}{2\tau}(z-z_o)^2
\end{eqnarray}
The non-translational invariant term $(x_oy-y_ox)$ in the action gives the Bohm-Aharonov phase, if we assume a straight line trajectory. Hence the
four-dimensional generalization will be
\begin{eqnarray}
K(x^\mu,\tau,x^{\prime\mu},0)=\Phi(x,x^\prime)\left(\frac{1}{4\pi\tau}\right)^2\frac{z}{\sin z}\nonumber\\
\exp\left(igH\cot z\frac{(\vec{x}_\perp-\vec{x}^\prime_\perp)^2}{4}\right) \exp\left(-i\frac{(x_\parallel-x_\parallel^\prime)^2}{4\tau}\right)
\end{eqnarray}
where $z=gH\tau$. It is now possible to write the propagator $K$ in momentum space. We use the Fourier transform
\begin{eqnarray}
\frac{z\pi}{i\tau\sin z}\exp\left(igH\cot z\frac{(\vec{x}_\perp-\vec{x}^\prime_\perp)^2}{4}\right)=\nonumber\\
 \int\frac{d^2p_\perp}{\cos z}\exp\left(-i\vec{p}_\perp\cdot(\vec{x}_\perp-\vec{x}_\perp^\prime)\right)\exp\left(-i\tau \frac{\tan z}{z}p_\perp^2\right)
\end{eqnarray}
This is similarly done for the longitudinal part ($\parallel$) of the propagator. Hence
\begin{eqnarray}
K(x^\mu,\tau,x^{\prime\mu},0)=\langle x|\exp\left(i\tau \Pi^2\right)|x^\prime\rangle\nonumber\\
=\Phi(x,x^\prime)\int\frac{d^4P}{(2\pi)^4}e^{iP\cdot(x-x^\prime)}\frac{e^{i\tau p_\parallel^2}}{\cos z}e^{-i\tau\frac{\tan z}{z}p_\perp^2}
\end{eqnarray}
where we have set $m=1/2$ and $\hbar=1$ in the nonrelativistic formula. This Formula is the same as
equation~(\ref{eq: tsai}) with appropriate relabeling of field and time parameters.
\bibliographystyle{unsrt}

\end{document}